\documentstyle[twocolumn,prl,aps,epsfig]{revtex}
\begin{document}
\draft
\twocolumn[\hsize\textwidth\columnwidth\hsize\csname @twocolumnfalse\endcsname
%
%
%

\title{Low-Temperature Scaling Regime of Random
  Ferromagnetic-Antiferromagnetic Spin Chains}

\author{ Beat Frischmuth and Manfred Sigrist}

\address{Institute of Theoretical Physics, ETH H\"onggerberg, CH-8093 Z\"urich,
Switzerland}

\date{\today}
\maketitle

\begin{abstract}
Using the Continuous Time Quantum Monte Carlo Loop algorithm, we
calculate the temperature dependence of the uniform susceptibility, 
and the specific heat of a spin-1/2 chain with random antiferromagnetic 
and ferromagnetic couplings,
down to very low temperatures. Our data show a consistent scaling
behavior in both quantities and support strongly the conjecture drawn
from the approximative real-space renormalization group treatment.
A statistical analysis scheme is developed which will be useful 
for the search scaling behavior in numerical and
experimental data of random spin chains.
\end{abstract}

\vskip2pc]
\narrowtext

Spin chains have for many decades attracted much interest  in
condensed matter physics and statistical mechanics. In recent years
random spin chains have been investigated intensively because
they provide a simple model to study the
interplay between quantum effects and disorder. There are
various realizations of quasi one-dimensional random spin systems
in nature. One class is represented by compounds like $Sr_3 CuPt_{1-x}Ir_x O_6$
\cite{Nguyen}. While the pure compounds $Sr_3 CuPtO_6$ $(x=0)$ and
$Sr_3 CuIrO_6$ $(x=1)$ are antiferromagnetic (AF) and ferromagnetic
(FM), respectively, the alloy $Sr_3 CuPt_{1-x}Ir_x O_6$ contains both
AF and FM couplings whose fraction is simply related to the
concentration $ x $ of $ Ir $. A corresponding minimal model is
the spin-$1/2$ Heisenberg chain, where the nearest neighbor
exchange coupling is $+J$ or $-J$ with
certain probabilities \cite{Furusaki}. Another related
example is realized in the low-temperature
regime of randomly depleted AF spin-1/2 Heisenberg
ladders. Two-leg Heisenberg ladders have a resonating valence bond
ground state which has short-range singlet correlation and a
spin excitation gap \cite{Dagotto}. If spins are depleted at random
the low-energy properties are changed drastically. Each site which
lost a spin is 
accompanied by an effective spin 1/2, and the residual interaction 
among these spins is randomly FM or AF with a wide distribution of 
coupling strengths \cite{Fukuyama,Sigrist}.

In this letter we focus on random FM-AF spin chains
with the full spin rotation symmetry. 
One of the most powerful techniques to study such systems is
the real space renormalization group (RSRG) method
\cite{Dasgupta,Ma,Hirsch1,Hirsch2,Fisher,West1,West}. This method was 
introduced to study random AF spin-$1/2$
chains \cite{Dasgupta,Ma,Hirsch1,Hirsch2,Fisher}, and was recently adapted 
to the study of the class of systems introduced above \cite{West1,West}. 
The basic idea of the RSRG method is the iterative 
decimation of degrees of freedom by integrating out successively the
strongest bonds in the spin chain. In this way the distribution of
coupling strengths and spin sizes is renormalized. For many cases 
a universal fixed point distribution 
is reached in the low-energy limit. 
The random AF spin-$1/2$ chain belongs to a universality class different from
that of the random FM-AF spin chain. The former is characterized by a 
{\it random singlet phase} where each spin tends to form a singlet with one
other spin even over very large distances. For
the FM-AF spin chain, however, spins correlate to form effective spins
whose average size 
grows with lowering of the energy scale 
\cite{West1,West,Hida}. A result of
the RSRG method is that the average number of original spins included in a
single effective spin for a given (energy or) temperature $ T $ scales
as $ \bar{l} \propto T^{- 2 \alpha} $, and the average spin size $
\bar{S} \propto T^{-\alpha} $ for $ T \to 0 $
\cite{West1,West}. In the latter case this RSRG
treatment gives a consistent description of the random spin chain, but
relies on certain approximations which have not so far been
independently tested.

Here we use a different approach to investigate the random FM-AF
spin-1/2 chain and its scaling behavior. Applying the recently-developed  
Continuous Time Quantum Monte Carlo (CT QMC) loop algorithm
\cite{Beard}, we calculate the temperature dependence of the uniform 
susceptibility and the specific
heat. The powerful CT QMC algorithm enables us to simulate 
systems down to very low temperatures, and thus to observe the
signs of the low-energy scaling regime of the fixed point. 
We find excellent
agreement between our analysis and the RSRG treatment by Westerberg et
al. \cite{West1,West}. Beyond the consistent evaluation of the scaling
exponent $ \alpha $, our simulation gives various quantities over a
wide range of temperatures and our discussion provides a technique to
analyze (numerical or experimental) data for the low-temperature
scaling regime.

For the numerical simulation we use as the model of a random
FM-AF spin-$1/2$ chain
\begin{equation}
H=\sum_{i}J_i\vec{S}_i \vec{S}_{i+1},
\end{equation}
where $\vec{S}_i$ is the $i^{th}$ spin of the chain and $J_i$, the
coupling strength of the $i^{th}$ bond, takes random values of both
positive and negative sign. We assume
the bond distribution 
\begin{equation}\label{dist} 
P(J)=\left\{ \begin{array}{ll}
                \frac{1}{2 J_0} & \mbox{$-J_0< J < J_0$} \\
                0               & \mbox{\rm otherwise,}
             \end{array}
     \right.
\end{equation}
where $J_0$ is the maximal coupling setting the energy scale.

According to Westerberg et al. \cite{West}, the low energy behavior of
the random FM-AF chain is independent of the initial distribution as
long as $J^{y_c} P(J)$ is regular for $J\rightarrow 0$ with $y_c
\approx 0.7$. Therefore, 
assuming a particular distribution (Eq. (\ref{dist})) is not a
restriction and our results apply qualitatively for all random spin
chains of this class. 

Using the CT QMC method, we simulate 100 samples of spin chains of 400
sites in a 
temperature range down to temperatures as low as
$T=J_0/1000$. In
the CT QMC loop algorithm the calculations are performed directly with a
Trotter time interval $\Delta\tau=0$, so no extrapolation in the
Trotter time interval is necessary. 
To obtain good statistics, we consider 100 different samples
of the spin chain following the distribution in Eq. (\ref{dist}).
The temperature dependences of the physical properties are calculated
for each chain separately, and then averaged over the 100 samples.  

An important result of the RSRG approach is that the uniform
susceptibility approaches the Curie-like behavior in the limit $ T \to
0 $. It is possible to determine the limiting Curie constant from
the initial spin size distribution and the ratio of FM and
AF bonds. The low-temperature deviations from this Curie constant 
give important information on the scaling
properties mentioned above. For the scaling analysis we calculate the
product $ T 
\chi(T) $ with the QMC simulation over a temperature range, $ 0 < T <
J_0 $ as shown in 
Fig. 1. According to the RSRG scheme with lowering temperature
gradually the spins correlate within clusters of growing length $ l
$. Each such cluster represents an effective spin. This leads in our case
to an effective diminishing of the spin degrees of freedom such that $
T \chi(T) $, as a measure for the still uncorrelated spins
at a given temperature, decreases
monotonically from the large-$T$ value of 1/4, the free spin-1/2
limit. The linear dependence in the intermediate temperature range
reflects the initial bond distribution $ P(J) $ in Eq. (\ref{dist}). 
This can be very clearly observed in the comparison of the QMC results
with the finite size 
systems, $L=2,3 $ and 4, for which we have performed the exact bond 
average (Fig. 1). The deviation from this linear behavior for $ T <
0.05 J_0 $ indicates the turn towards the low-temperature
scaling regime. Here the
effect of the renormalization of the distribution function becomes visible.
Note that such renormalization effects are already possible for finite
size systems with $ L= 4 $ as can be seen in Fig.1. 
In the low-temperature 
regime thermodynamic quantities should approach universal 
behavior which we will now analyze based on the scaling assumption of the
RSRG method. 

\begin{figure}
\epsfxsize=85mm
\epsffile{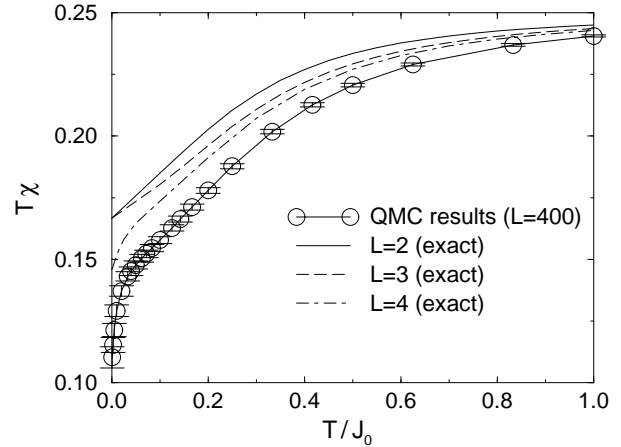}
\caption[*]{Temperature dependence of the uniform Curie constant of
  random chains of different length $L$ with a bond distribution
  according to (\ref{dist}). The data for chains of two, three and
  four sites are calculated exactly, integrating over the distribution
  (\ref{dist}) for each bond. The data for $L=400$ are QMC results.} 
\end{figure}

The fact that the correlation of spins occurs within definite clusters
permits the use of a special statistical approach for the discussion of the
thermodynamic 
properties. The effective spin of a cluster of length $ l $ is equal
to the ground state spin quantum number. For spin-1/2 degrees of
freedom coupled randomly by FM or AF bonds the probability
distribution for these quantum numbers is given by,

\begin{equation}
\rho_{l}(S) = \frac{l!}{(l/2+S)! (l/2-S)!}
                \frac{1}{2^l} \\ (2 - \delta_{S,0}),
\end{equation}
where $ S $ ranges from 0 to $ l /2 $ ($S$: integer for $l$ even 
and half-integer for $l$ odd). Because the system contains no 
frustrating couplings, the spin quantum number $ S $ is
identical  to the total spin of the corresponding 
completely correlated cluster of classical spins \cite{Sigrist,West1,West}.

The average value of the Curie constant per site for a finite, 
completely correlated cluster of length $l$ is
\begin{equation}\label{curie1}
\mbox{$\mathcal{C}$}_l = \frac{1}{3l} \sum_{S=0}^{l/2} \rho_{l}(S)
S(S+1).
\end{equation}
Now let us consider the distribution of cluster lengths $ l $ for a
given temperature. It is natural to assume that the bonds freeze 
uncorrelated among each other, i.e. each bond  
is frozen with a certain probability independent of the location
within the chain. Thus
the distribution of $ l $, $ p_{\bar{l}}(l) $ has an exponential form. 
As mentioned above, the RSRG approach predicts a scaling behavior for the 
temperature dependence of the average cluster length size $\bar{l}$. 
\begin{equation}\label{length}
\bar{l}=\lambda^2\left(\frac{T}{J_0}\right)^{-2 \alpha},
\end{equation}
where $\lambda$ is a dimensionless proportionality factor. 
This can now be used to calculate the low-temperature behavior of
$ T \chi $. If $ \bar{l} \gg 1 $, it is justified that we use
the distribution functions $ \rho_l (S) $ and $ p_{\bar{l}}(l) $ in
their continuum approximations 

\begin{equation} \begin{array}{l} \label{continuum}
\rho_l (S) = 2 \sqrt{\frac{2}{\pi l}} e^{-2S^2/l} \\ \\ 
p_{\bar{l}}(l)= \frac{e^{1/(\bar{l} -1)}}{\bar{l} -1} e^{-l/(\bar{l}
  -1)} \\
\end{array} \end{equation}
for $ l \geq 1 $. (We have compared the following results also
numerically with the discrete  
forms and found only minor deviations even for small $ \bar{l}$.) The 
averaging over both distributions leads to

\begin{eqnarray} \label{curie2}
T \chi(T) & = & \frac{1}{\bar{l}} \int_1^{\infty} dl\,p_{\bar{l}} (l) l 
\mbox{$\mathcal{C}$}_l \nonumber \\
        &= &\frac{1}{12} + \frac{\bar{l}^{-1/2}}{6 \sqrt{2}} + 
		O(\bar{l}^{\,-3/2})
\nonumber \\ 
        &= & \frac{1}{12} + \frac{T^{\alpha}}{6 \lambda \sqrt{2}} + O(T^{3
  \alpha}), 
\end{eqnarray}
where for the third line $\bar{l}$ was substituted by
Eq. (\ref{length}). Here the idea enters that the effective
spins formed by each cluster behave independently under
thermal fluctuations.

Fitting the leading two terms of the above form (Eq. (\ref{curie2}))
to the QMC data gives an estimate of the scaling exponent $\alpha$
and the proportionality factor $\lambda$. Fig. 2 shows $T\chi$ as a
function of $T^\alpha$ for different $\alpha$. Using the
\begin{figure}
\epsfxsize=85mm
\epsffile{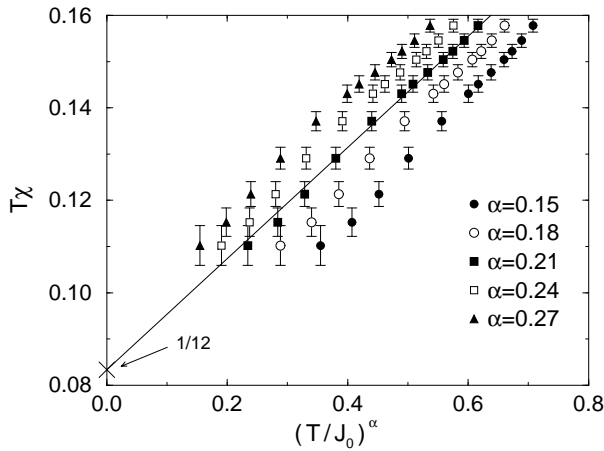}
\caption[*]{Uniform Curie constant per spin as a function of
  $T^\alpha$ for different $\alpha$. The best linear behavior
  (Eq. (\ref{curie2})) is found for $\alpha=0.21\pm0.02$. The solid
  line shows a fit of form (\ref{curie2}) for $\alpha=0.21$ and
  fixed intercept $1/12$. This leads to a value of $\lambda=1.0\pm
  0.1$ .} 
\end{figure}

\begin{figure}
\epsfxsize=85mm
\epsffile{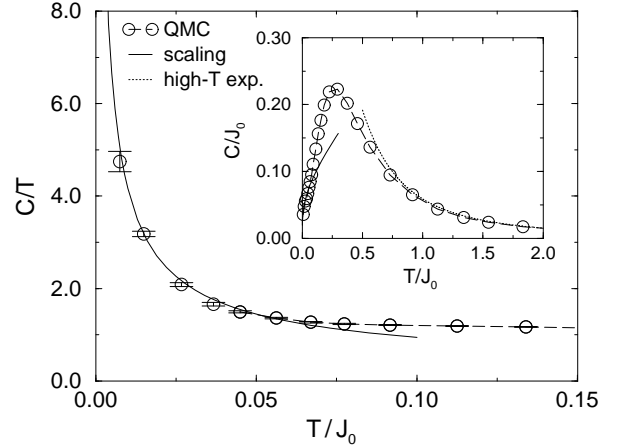}
\caption[*]{Specific heat divided by temperature $T$ as a function
  of $T$. The circles are the QMC results, while the solid line was
  calculated as the derivative of Eq. (\ref{entropy}) with respect to
  $T$. The inset shows the QMC results for the specific heat as a
  function of temperature over a large temperature range. The
  error bars are smaller than the symbols. } 
\end{figure}

\hspace*{-0.55cm} knowledge of the exact zero-temperature value of 
$ T \chi $ ($ = 1/12 $) we obtain from this
plot $\alpha=0.21 \pm 0.02$, in good agreement with
the RSRG result ($ 0.22 \pm 0.01 $) \cite{West}. The
proportionality factor is determined as $\lambda=1.0 \pm
0.1$. These results confirm that our QMC
results for chains of length $L=400$ are not affected by finite-size
effects, because according to Eq. (\ref{length}) the average length of
the correlated clusters, even at the lowest simulated temperature $J_0
/1000$, is only $\bar{l}\approx 18$, and therefore much smaller than
$L$.  

Now we turn to  the specific heat $C$, which is
determined from the numerical derivative of the
internal energy  calculated in the QMC simulations. 
The errors in the specific heat increase significantly 
with lowering temperature. Nevertheless, $C$ could be determined 
reliably down to $T\approx J_0/150$. The QMC results 
for the specific heat and the
ratio $C/T$ as a function of $T$ are shown in Fig. 3.  

Since $C/T$ is the derivative of the entropy $\sigma$ with respect to
$T$, we may compare it to a calculation of $\sigma$ by our statistical
 cluster analysis. Again we assume that for given temperature each
 correlated cluster behaves independently. Hence a cluster of length $ l
 $ contributes 
\begin{equation}\label{entropy1}
\sigma_l=\sum_{S=0}^{l/2} \rho_l(S) \log(2S+1).
\end{equation}
to the entropy. 
Here it is necessary to use the discrete expressions for the 
distribution functions because large deviations arise in the 
continuum approach using Eq. (\ref{continuum}). The discrete 
distribution $ \tilde{p}_{\bar{l}} (l) $ now has the form

\begin{equation}\label{distpdis}
\tilde{p}_{\bar{l}}(l)=(e^\kappa-1) e^{-\kappa l}, \ \ \mbox{\rm with} \
\kappa=\log (1+\frac{1}{\bar{l}-1}). 
\end{equation}
Note that this distribution is normalized for a sum over $ l $ from 1
to $ \infty$. The entropy per site of the infinite chain is
\begin{equation}\label{entropy}
\sigma(T)=\frac{1}{\bar{l}}\sum_{l=1}^\infty p_{\bar{l}}(l)\, \sigma_l ,
\end{equation}
where $\bar{l}$ is given by Eq. (\ref{length}). 
Using the values of $\alpha$ and $\lambda$ determined above,
the entropy and hence $C/T$ may be calculated from
Eq. (\ref{entropy}). The result is shown as the solid line 
in Fig. 3. We find good agreement with the QMC results for the 
low-temperature range and emphasize that no further fitting was 
required to obtain this result. At higher temperatures this approach
must fail, because the scaling behavior of $ \bar{l} $
(Eq. (\ref{length})) is no longer valid.

As a further test of the consistency of our analysis, we estimate
the area below 
the curve $C/T$ from $T=0$ to $\infty$ using 
$\sigma(\infty)=\sigma(T^*)+\int_{T^*}^\infty dT \, (C/T)$, 
where we set $T^* =0.03 J_0$. The first term is obtained
using Eq. (\ref{entropy}) and the integral from $T^*$ to $ \infty $ 
is determined numerically, using the QMC results and a
high temperature expansion up to third order. We find 
$\sigma(\infty)=0.68 \pm 0.01$, in very good agreement with the 
expected result $\ln 2$. The fraction $\sigma(T^*)$ of the entropy 
is large, contributing approximately 25\% of the total. Therefore the fact 
that we find the correct value for $\sigma(\infty)$ is a convincing 
test for the correctness of Eq. (\ref{entropy}), and of our statistical 
treatment of the scaling regime.

In the study of the random FM-AF spin-$1/2$ chain, the powerful
CT QMC loop algorithm allowed us to reach temperatures where clear
characteristics of the universal low-energy properties of this system are
observable. The characteristic temperature where the
scaling behavior begins depends on the initial distribution
of bond strengths. Our choice of a non-singular distribution has
the advantage that this temperature is not so low. The key
feature for analyzing the low-temperature data lies in the fact that 
the spins are correlated within clusters whose length grows with
decreasing temperature. The scaling of the average length 
$\bar{l}=\lambda^2 (T/J_0)^{-2\alpha}$ and the statistics of the
effective spin distribution allow a very good analysis of the numerical data.
We determine both the prefactor and the exponent. The latter is
assumed to be universal and agrees surprisingly well with the value found by
RSRG \cite{West1,West}. 
Among the (unfrustrated) random spin chains with complete spin
rotation symmetry there are only two distinct classes, the ones with
the random singlet phase fixed point and the others which belong to
the class studied here \cite{singular}. For the former it was shown by
Fisher that the RSRG scheme describes the fixed point behavior exactly
\cite{Fisher}. This was not possible so far for the random FM-AF spin
chain. Our numerical and statistical analysis, however, demonstrates very
convincingly  the validity of the universal scaling
assumption. Finally, we
would like to emphasize that our statistical fitting
procedure is very suitable and useful to analyze not only numerical,
but also experimental results
of this class of random spin systems.

We would like to thank A. Furusaki, P.A. Lee, N. Nagaosa, B. Normand,
T.M. Rice, E. Westerberg for many fruitful discussions, and especially
M. Troyer and B. Ammon for their assistance concerning the QMC algorithm. 
We are also grateful for financial support from the Swiss Nationalfonds. 
In particular, M.S. is supported by a PROFIL-Fellowship. The calculations 
were performed on the Intel Paragon at the ETH Z\"urich.


\begin{thebibliography}{4}
\bibitem{Nguyen}  T.N. Nguyen, P.A. Lee, and H.-C. zur Loye, Science \bf 271\rm , 489 (1996).
\bibitem{Furusaki} A. Furusaki, M. Sigrist, P.A. Lee, K. Tanaka, and N. Nagaosa, Phys. Rev. Lett \bf 73 \rm (1994) 2622.
\bibitem{Dagotto} For a review, see E. Dagotto and T.M. Rice, Science \bf 271 \rm (1996) 618
\bibitem{Fukuyama} H. Fukuyama, T. Tanimoto and M. Saito, J. Phys. Soc. Jpn. \bf 65 \rm (1996) 1183; H. Fukuyama, N. Nagaosa, M. Saito and T. Tanimoto, J. Phys. Soc. Jpn. \bf 65 \rm (1996) 2377.
\bibitem{Sigrist} M. Sigrist and A. Furusaki, J. Phys. Soc. Jpn. \bf 65 \rm  (1996) 2385.
\bibitem{Dasgupta} C. Dasgupta and S.-k. Ma, Phys. Rev. B \bf 22 \rm (1980) 1305.
\bibitem{Ma} S.-k. Ma, C. Dasgupta, and C.K. Hu, Phys. Rev. Lett. \bf 43 \rm (1979) 1434.
\bibitem{Hirsch1} J.E. Hirsch and J.V. Jose, Phys. Rev. B \bf 22 \rm (1980) 5339.
\bibitem{Hirsch2} J.E. Hirsch, Phys. Rev. B \bf 22 \rm (1980) 5355.
\bibitem{Fisher} D.S. Fisher, Phys. Rev. B \bf 50 \rm (1994) 3799.
\bibitem{West1} E. Westerberg, A. Furusaki, M. Sigrist and P.A. Lee, Phys. Rev. Lett. \bf 75 \rm (1995) 4302.
\bibitem{West} E. Westerberg, A. Furusaki, M. Sigrist and P.A. Lee, cond-mat/9610156.
\bibitem{Hida} K.Hida, preprint.
\bibitem{Beard} B.B. Beard and U.-J. Wiese, cond-mat/9602164.
\bibitem{Nagaosa} N. Nagaosa, A. Furusaki, M. Sigrist, and
  H. Fukuyama, J. Phys. Soc. Jpn. \bf 65 \rm (1996) 3724.
\bibitem{singular} Random spin
  chains whose bond distribution functions are very singular in the
  limit $ J \to 0 $ are excluded from this classification. For these
  systems, usually no universal scaling 
  regime can be reached and the low-temperature behavior is governed
  by the $ J \to 0 $-limit of the initial bond distribution.
\end{thebibliography}
\end{document}